\documentclass[superscriptaddress,twocolumn]{revtex4}
\usepackage{graphicx,color,url,ulem}
\usepackage{amsmath}
\usepackage{amssymb}

\begin{document}

\newcommand{\MDB}{Institute of Physics, Otto-von-Guericke-University, D-39106 Magdeburg, Germany}

\newcommand{\SZFI}{Institute for Solid State Physics and Optics, Wigner Research Centre
for Physics, Hungarian Academy of Sciences, P.O. Box 49, H-1525 Budapest, Hungary}

\newcommand{\CAIRO}{Faculty of Engineering and Technology, Future University in Egypt, New Cairo, Egypt}

\title{Flow of anisometric particles in a quasi-2D hopper}

\author{Bal\'azs Szab\'o}

\affiliation{\SZFI}

\author{Zsolt Kov\'acs}
\affiliation{\SZFI}

\author{Sandra Wegner}
\affiliation{\MDB}

\author{Ahmed Ashour}
\affiliation{\MDB}, \affiliation{\CAIRO}

\author{David Fischer}
\affiliation{\MDB}

\author{Ralf Stannarius}
\affiliation{\MDB}

\author{Tam\'as B\"orzs\"onyi}
\email{borzsonyi.tamas@wigner.mta.hu}
\affiliation{\SZFI}

\begin{abstract}
The stationary flow field in a quasi-two-dimensional hopper is investigated experimentally.
The behavior of materials consisting of beads and elongated particles with different aspect ratio
is compared. We show, that while the vertical velocity in the flowing region can be fitted with
a Gaussian function for beads, in the case of elongated grains the flowing channel is narrower
and is bordered with sharper velocity gradient. For this case, we quantify deviations from the
Gaussian velocity profile. Relative velocity fluctuations are considerably larger and slower
for elongated grains.
\end{abstract}

\maketitle

\section{Introduction}
\label{intro}

Hopper flows are very important in agriculture and various industrial
processes dealing with granulates. The basic features of such flows have
been characterized in numerous experimental and numerical studies for spherical grains, and
more recently increasing attention has been payed to systems involving nonspherical particles.
One of the fundamental questions relates to the outflow rate as a function of the orifice size,
for which a power law behavior was found by Beverloo et al.~\cite{beverloo1961}.
This relationship has been tested for various materials and was finetuned
for the small particle-to-outlet diameter ratio limit \cite{mankoc2007}.
Comparing the flow rate of spherical and slightly elongated particles (with equal volume)
numerical (DEM) studies predicted \cite{cleary1999,cleary2002,liu2014} that for frictional
grains increasing particle elongation leads to lower flow rates, which was recently confirmed
by experimental investigations \cite{ashour2017}.
On one hand, this might be counterintuitive, as elongated grains undergo shear induced
orientational ordering with their average orientation pointing almost in the direction of
the flow lines \cite{baxter1990,ankireddy2009,campbell2011,borzsonyi2016,borzsonyi2012}.
On the other hand, several authors have shown, that for other nonspherical grains increasing
grain angularity (or in other words increasing effective friction of the material) reduces
the mass flow rate \cite{hohner2012,hohner2013,cleary1999,cleary2002,sukumaran2003,soltanbeigi2017},
and leads to larger stagnant zones
and more residual mass after discharge \cite{gupta2012,hohner2012,hohner2013,wu2008}.
Several authors detected and quantified fluctuations in the discharge rate or flow field
\cite{medina1998,baxter1989,mollon2013,unac2012}. The amplitude of the relative fluctuations
of the discharge rate \cite{unac2012,janda2009} or flow velocity \cite{thomas2016} was shown
to increase with decreasing orifice size,
and finally, the probability for clogging \cite{zuriguel2005,zuriguel2014} appears to
increase with increasing particle aspect ratio  (length $L$ to diameter $d$) \cite{ashour2017}.

The flow field inside a hopper can be approximated from microscopic arguments. In the Void Model
of Litwiniszyn and Mullins \cite{litwiniszyn1963,mullins1972} particles move downward by falling
into holes below them, thus the flow is related to directed (upward) random walks of particle
sized voids from the orifice. This leads to a Gaussian velocity profile across the hopper
as it was elaborated in the Kinematic Model of Nedderman and T\"uz\"un \cite{nedderman1979}.
Even if the diffusive nature and the Gaussian profile was experimentally confirmed by
several groups using beads \cite{medina1998,choi2005,garcimartin2011}, particle tracking
or DEM simulations did not confirm the simple microscopic mechanism described above
\cite{choi2004,arevalo2007}.
Namely, in the diffusion equation, the diffusion constant was shown to depend on the distance
from the orifice \cite{medina1998,choi2005,balevicius2011}.
Bazant and Rycroft showed, that considering the collective rearrangement of a spot of
grains ("\textit{spot model}") better microscopic agreement is observed, and the introduction
of a new length scale (spot size) helped to resolve some of the discrepancies
\cite{bazant2006,rycroft2006}. This idea was further elaborated as a "stochastic flow rule"
\cite{kamrin2007,kamrin2007-2} or nonlinear elasto-plastic model \cite{kamrin2010}, capable to
describe flowing regions and stagnant zones in granular flows simultaneously.
Another recent numerical work by Staron et al.~showed, that the velocity profiles obtained
by a discrete Contact Dynamics algorithm are reproduced when the $\mu(I)$ flow law
(obtained experimentally for glass beads \cite{jop2005,jop2006}) is incorporated into a continuum
Navier-Stokes solver \cite{staron2014}.

The flow field for nonspherical grains is less investigated.
Pioneering Particle Image Velocimetry (PIV) measurements were carried out with cylinders of
aspect ratio close to 1
and slightly nonspherical (Amaranth) grains \cite{steingart2005,sielamowicz2005}, but
the given sample velocity profiles were not fitted by any function.
Ellipses with an aspect ratio of $1.3$ were also tested in a two dimensional silo \cite{favier2001}.
Although the experimental data were quite noisy, they were fitted using a Gaussian function.
More recently, the velocity profile for Amaranth grains was found to be closer to a parabolic
function than to a Gaussian \cite{sielamowicz2011}.
Discrete element simulations with corn shaped particles reported slightly larger grain velocity
in the center of the hopper compared to the case of beads, but no further analysis of the velocity
profiles was presented \cite{tao2010,gonzales2012}.

In the present work we determine the velocity fields by PIV analysis for glass rods with two different
aspect ratios ($L/d=1.4$ and $L/d=3.5$), plastic rods with $L/d=6$, lentils (aspect ratio of 0.4), and two type of beads
(silica gel and plastic).
We observe and quantify deviations from the Gaussian velocity profile for rods.
We show, that the amplitude of temporal fluctuations of the velocity field systematically increases
with particle elongation.


\section{Experimental methods}
\label{exp}

\begin{figure}[!ht]
\includegraphics[width=0.96\columnwidth]{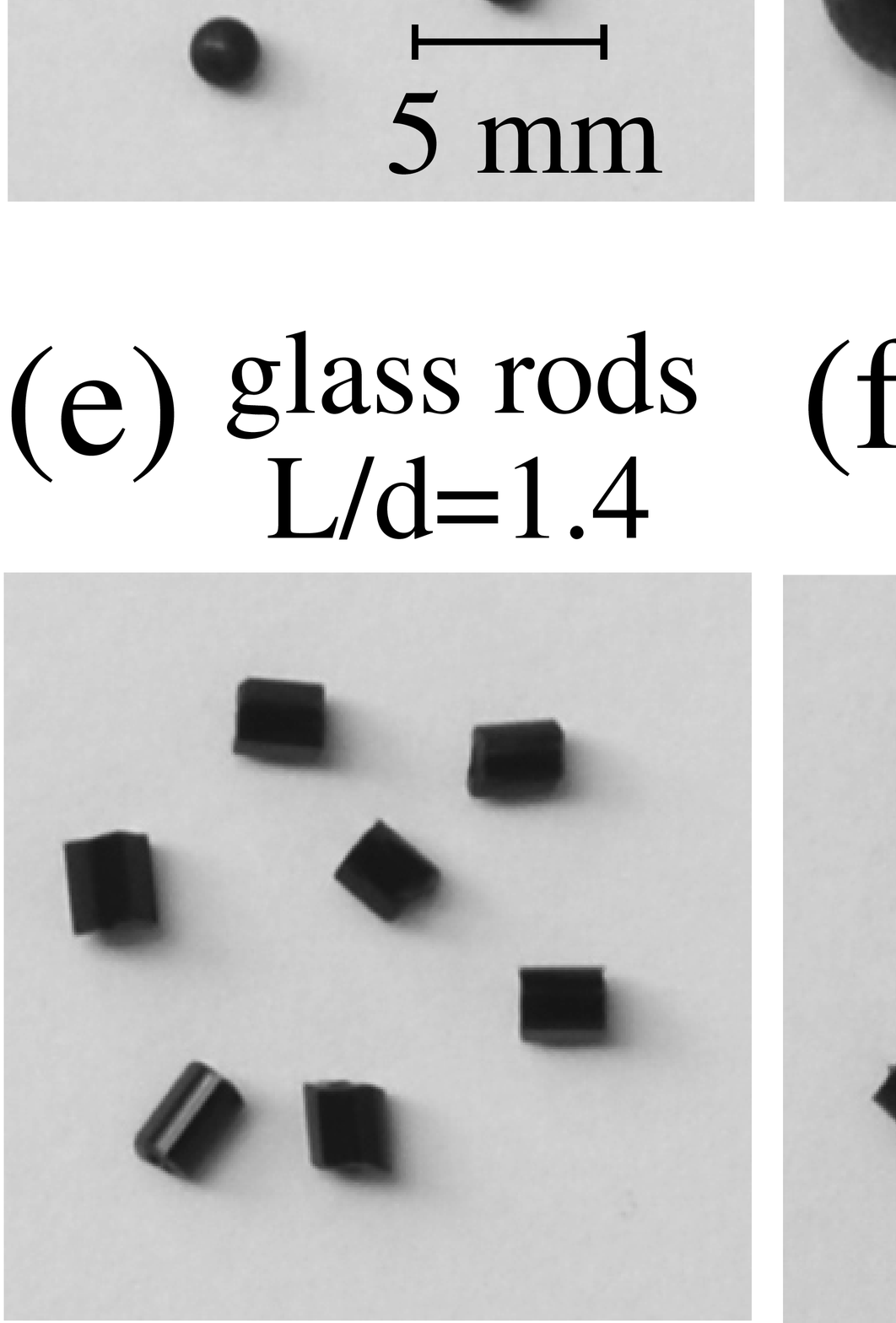}
\caption{(a) Schematic view of the experimental geometry. Dashed line indicates the observation area.
(b-f) Photographs of the granular samples.
}
\label{fig:setup}
\end{figure}

In our experiments a quasi-two-dimensional hopper was used (see Fig.~\ref{fig:setup}(a)), with
horizontal and vertical dimensions of $700$ and $600~\mathrm{mm}$, respectively.
%
The central area ($290~\mathrm{mm} \times 370~\mathrm{mm}$) was recorded by a digital camera
(\textit{MotionBLITZ EoSens mini}, $1.2$ MPixel with a frame rate of $200$ fps).
The distance between the two glass plates $W$ was set to $18~\mathrm{mm}$ (similar results were
obtained with $W=35~\mathrm{mm}$).
The additional reservoir at the top of the hopper helped reducing finite size effects, i.e. ensures
that the flow field is not influenced by surface distortions of the quasi 2D granular layer.
In the measurements presented here we used $4$ different orifice sizes $D$ ($15 - 45~\mathrm{mm}$)
and $7$ different values ($46^{\circ} - 140^{\circ}$) of the inclination angle $\Phi$ of the wedge
shaped walls. The flow field, which is essentially restricted to a two-dimensional plane, 
was detected by a self-written PIV algorithm, which is basically similar to other freely or 
commercially available PIV codes. Focusing on vertical motion,
the box size for correlating segments of the subsequent images was chosen to be much larger
in the $z$ direction than the $x$ direction. This allowed us to determine the vertical displacement
of the image segments with high (subpixel) resolution, low noise, and with increased horizontal
resolution of the data points.
This algorithm was used to determine high frequency oscillations in a cylindrical hopper flow
\cite{borzsonyi2011}, or high resolution displacement profiles in sheared granular media \cite{szabo2014}.
In the present study, $3$ measurements were recorded for each setting, and the results were
averaged to reduce statistical fluctuations.
Photographs of the $six$ granular samples are shown in Fig.~\ref{fig:setup}:
spherical silica gel beads ($d=1.8~\mathrm{mm}$, Fig.~\ref{fig:setup}(b)),
airsoft balls ($d=6.0~\mathrm{mm}$, Fig.~\ref{fig:setup}(c)), and oblate
lentil seeds ($L=2.5~\mathrm{mm}, d=6.4~\mathrm{mm}, L/d = 0.4$, Fig.~\ref{fig:setup}(d)),
short glass rods ($L=2.5~\mathrm{mm}, d=1.8~\mathrm{mm}, L/d=1.4$, Fig.~\ref{fig:setup}(e)),
long glass rods ($L=6.6~\mathrm{mm}, d=1.9~\mathrm{mm}, L/d=3.5$, Fig.~\ref{fig:setup}(f)),
and plastic rods ($L=14~\mathrm{mm}, d=2.33~\mathrm{mm}, L/d=6.0$, Fig.~\ref{fig:setup}(g)).
The choice of materials allows us to investigate new types of beads (silica gel and plastic),
to complement earlier measurements on glass beads \cite{medina1998,choi2005} and steel beads
\cite{garcimartin2011}, and to study the case of nonspherical particles with similar size as
the beads.

In the experimental procedure the cell was filled first by closing the outlet and pouring the
granulates from above. When the flow was started, an initial transient occurred, during which
the width of the flow was continuously decreasing.  After a few seconds, the flow profile became
stationary. Near the end of the run, the free surface of the granular layer approached the
observation area, and the flow profile became wider again.
We focus on the stationary flow between the initial and final transients.

\begin{figure}[!ht]
\includegraphics[width=0.99\columnwidth]{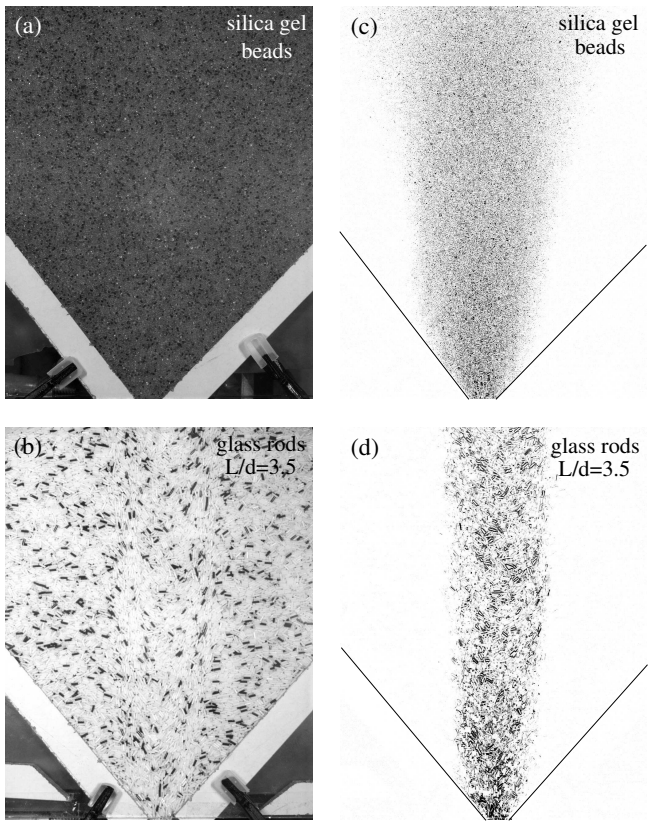}
\caption{(a) Sample images of the flow for (a) silica gel beads, and (b) glass rods with $L/d=3.5$.
(c)-(d) The flowing region visualized by image differencing.
}
\label{fig:images}
\end{figure}

Sample images taken during stationary flow are presented for silica gel beads and for glass rods with
$L/d=3.5$ in Figs.~\ref{fig:images}(a)-(b)). Visualizing the moving regions (Figs.~\ref{fig:images}(c)-(d))
%
by taking the difference of subsequent images shows that flow is concentrated in a narrower channel
(i.e. the stagnant zone is larger) for the case of rods.
A video showing the temporal evolution of the system can be found in the supplementary material,
where a third column is inserted showing the velocity field calculated by the PIV algorithm.


\section{Results}

\begin{figure}[!htbp]
\includegraphics[width=\columnwidth]{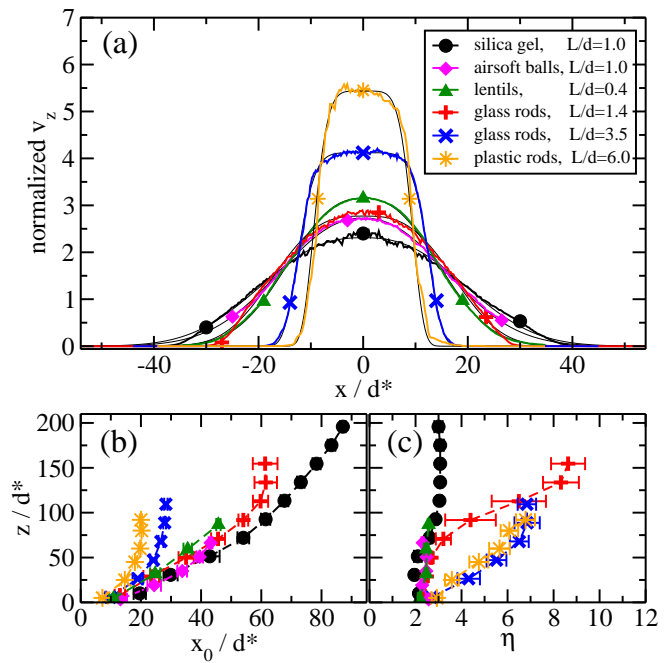}
\caption{(color online).
(a) Time averaged vertical velocity as a function of $x$, at height $z=60~d^*$, for all $5$ materials.
The curves are normalized by the integral of the fitted functions (Eq.~\ref{eq:fit-function}),
symbols are displayed on each curve for a better distinction.
The thin black lines represent the fitted curves.
(b)-(c) The two fitting parameters: the exponent $\eta$ and the half width of the flowing
channel $x_0$ (defined in Eq.~\ref{eq:fit-function}) at different heights. Both $x_0$ and the
height coordinate $z$ are in units of the effective particle diameter d*. For each material
the outlet size was $D \approx 7.5~d^*$, the angle of the wedge shaped walls was set to
$\Phi = 80^{\circ}$.
}
\label{fig:velocity}
\end{figure}

In order to quantitatively compare the flow fields for different materials, the time averaged
vertical velocity ($v_{z}(x)$) has been determined across the sample. These profiles are shown in
Fig.~\ref{fig:velocity}(a) at the height of $z=60~d^*$, for all six samples at the same dimensionless
orifice size $D \approx 7.5~d^*$. Here $d^*$ stands for the equivalent diameter of a sphere having
the same volume as the elongated particle.

%
The velocity curves have been fitted with the function:

\begin{equation}
        \label{eq:fit-function}
        v_{z}= c \  \text{exp}\left( { - \left(  \frac{2|x|}{x_0} \right) ^ {\eta} } \right)\quad ,
\end{equation}

where the exponent $\eta$ quantifies deviations from the Gaussian ($\eta=2$) velocity profile, and
$x_0$ is the half width of the flowing channel. Larger $\eta$ make the slopes of the profile steeper,
small $\eta<2$ smoothen the profile. These two parameters have been determined
as a function of the vertical coordinate $z$ and are shown in Fig.~\ref{fig:velocity}(b)-(c).
As seen, for $3$ samples the exponent $\eta$ stays around $2$ and the normalized value of the channel
width ($x_0/d^*$) is increasing with $z$ very similarly.
As mentioned above, the velocity profiles were shown to be Gaussian for spherical glass
beads \cite{medina1998,choi2005} and steel beads \cite{garcimartin2011} in earlier studies.
Our data for beads nicely confirm the appropriateness of Gaussian fits for two further types of
material: silica gel and plastic airsoft balls. The case of smooth oblate particles (lentils)
also appear to obey this rule.

For the three samples consisting of rods however, the exponent $\eta$
becomes significantly larger than $2$ above a certain value of $z$. At around the same height
the flow width $x_0/d^*$ starts deviating from the other $3$ curves. Thus for the case of rods
above a certain height, the the velocity profile is characterized by a plateau with relatively
narrow shear zones at the two sides. The height above which the exponent $\eta$ substantially
deviates from 2 depends on the grain shape,
and is about $20~d^*$ for rods with  $L/d=3.5$ and $L/d=6$ and around $100~d^*$ for rods with $L/d=1.4$.
This shows, that the velocity field for the longer rods ($L/d=3.5$ and $L/d=6$) is
non-Gaussian almost in the entire hopper,
while for $L/d=1.4$ a region right above the outlet remains Gaussian.
We note, that the largest value of the exponent $\eta$ is detected for these shorter
rods ($L/d=1.4$), we will get back to this observation later.

\begin{figure}[!ht]
\includegraphics[width=\columnwidth]{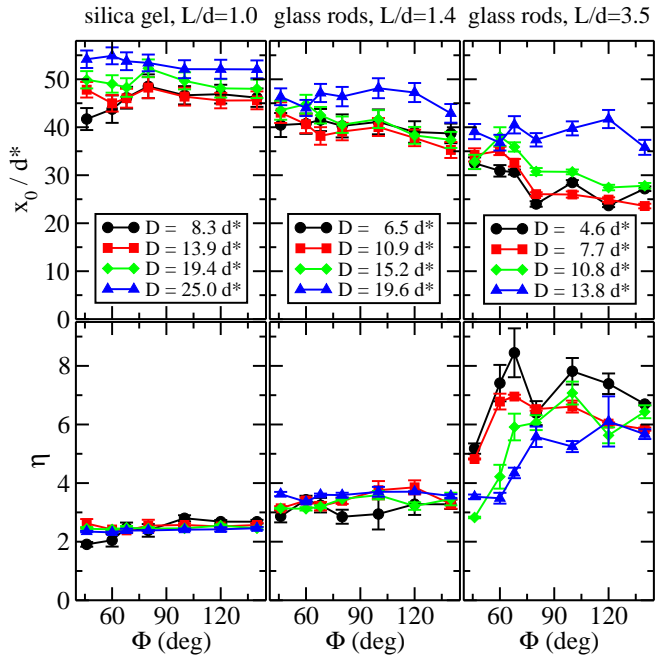}
\caption{(color online).
The two parameters describing the velocity profile: the exponent $\eta$ and the normalized
half width $x_0/d^*$ (see Eq.~\ref{eq:fit-function}) as a function of the hopper angle $\Phi$,
at a height of $z = 60~d^*$, for three of the investigated materials. The lines merely guide the eye.
}
\label{fig:flowwidth-angle-all}
\end{figure}

The systematic change in the exponent $\eta$ and the half width of the flowing channel
$x_0$ by changing the grain elongation is also demonstrated in Fig.~\ref{fig:flowwidth-angle-all}.
We see, that this behavior appears to be general, as both $\eta$ and $x_0$ do not change
significantly when changing the hopper angle $\Phi$ or the orifice size $D$.
%
%
Looking at the details, a slight increase of $x_0$ can be noted with increasing $D$ for all
three materials and with decreasing $\Phi$ for rods. In the experiments with longer glass
rods, deviations from the Gaussian profile are stronger (i.e. $\eta$ is larger) for larger hopper
angles $\Phi$.

\begin{figure}[!ht]
\includegraphics[width=\columnwidth]{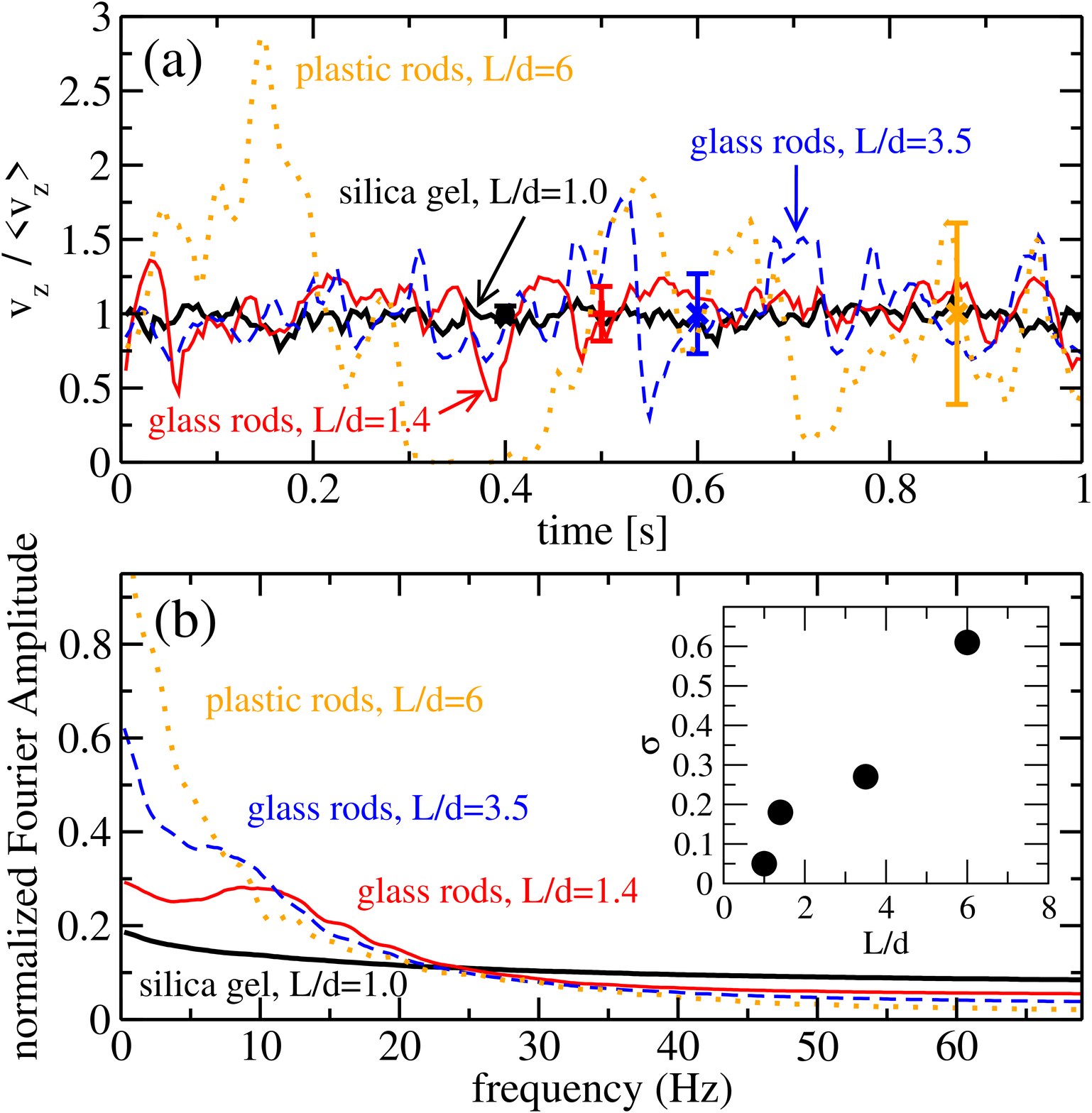}
\caption{(color online).
(a) Temporal fluctuations in the velocity values above the orifice at $z=60~d^*$.
(b) Fourier power spectrum of the curves presented in panel (a), normalized by their integral
across the whole frequency range (from $0$ to $100~\mathrm{Hz}$).
The inset shows the standard deviation of the velocity data as a function of the
particle elongation $L/d$.
}
\label{fig:fluctutions-flow-clogging}
\end{figure}

The movie presented as supplementary material visualizes temporal fluctuations of the flow
velocity. The time evolution of the velocity taken in the central part of the hopper at
$z=60~d^*$ is shown in Fig.~\ref{fig:fluctutions-flow-clogging}(a).
As it is seen, the amplitude of relative deviations from the mean velocity systematically
increases with grain elongation. This is quantified by the standard deviation $\sigma$
of the normalized velocity data which is shown as a function of $L/d$ in the inset of
Fig.~\ref{fig:fluctutions-flow-clogging}(b). We note that the actual time sequences are longer
(about 4 s), Fig.~\ref{fig:fluctutions-flow-clogging}(a) shows only a 1 s interval, so that
the timescale of the fluctuations is better seen. Another way to characterize the time sequences
is to measure the asymmetry of the fluctuations. This can be quantified by the fraction of velocity
data points above and below the average, which are distributed at 53:47 for silica gel, 56:44 
for short glass rods, 42:58 for long glass rods and 41:59 for plastic rods with $L/d=6$.
The asymmetry is increasing with grain elongation, and notably it changes sign for the samples with
longer grains which show larger velocity fluctuations. As described above, three independent runs
were performed for each material. The mean velocity calculated for these runs varied less than
 $1.5 \%$ for silica gel, $3 \%$ for rods with $L/d=1.4$, $12 \%$ for rods with $L/d=3.5$,
and $10 \%$ for rods with $L/d=6$. Thus even if the velocity fluctuations were relatively large,
the mean velocity measured in independent runs varied comparably little.

Performing a Fourier analysis of the $v_z(t)$ signals reveals,
that increasing grain elongation leads to increasing amplitude in the low frequency range
of the power spectrum (see Fig.~\ref{fig:fluctutions-flow-clogging}(b)).
For rods with $L/d=1.4$ and $L/d=3.5$ noticeable peaks are seen at around $\approx 10~\mathrm{Hz}$
and $\approx 7~\mathrm{Hz}$, respectively.
%
%
Thus, for longer grains the velocity field fluctuates with larger amplitude and lower frequency.
From this respect, it would be worth investigating fully 3D hoppers, where the orientation of
rods is not influenced by the confining walls.
In any case, the above described observation is coherent with our recent findings on 3D hoppers,
where an increasing aspect ratio of the grains lead to lower flow rates and higher clogging
probabilities compared to spherical grains \cite{ashour2017}.

\begin{figure}[!ht]
\includegraphics[width=\columnwidth]{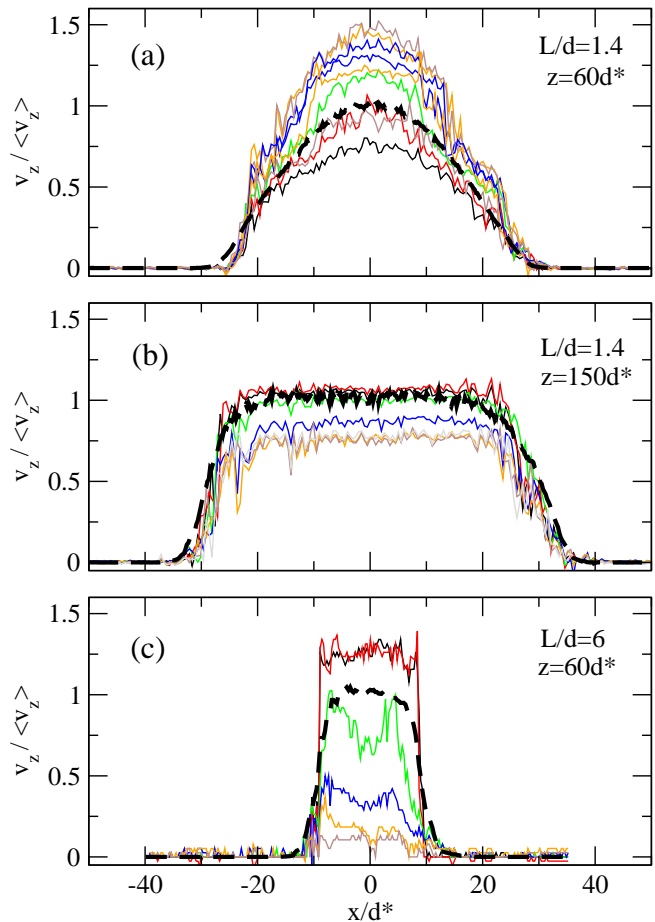}
\caption{(color online).
The form of velocity profiles during fluctuations for rods with $L/d=1.4$
at (a) $z=60~d^*$  and (b) $z=150~d^*$ and (c) rods with $L/d=6$ at $z=60~d^*$.
Different colors represent different instant velocity profiles, the dashed lines show the 
time averaged velocity profile for the whole run.
}
\label{fig:collapse}
\end{figure}

In the following, we analyze the effect of the fluctuations on the shape of the velocity profile.
Figure \ref{fig:collapse} presents velocity profiles taken from subsequent frames of the image sequence
from a selected period of time, when the velocity changes significantly.
Figure \ref{fig:collapse}(a)-(b) shows the case of rods with $L/d=1.4$ at the elevations of $z=60~d^*$
and $z=150~d^*$. As it is seen, the shape of the velocity profile remains similar, even if the
velocity value changes substantially. The time averaged velocity profile (shown with a dashed line)
is very similar to the instantaneous profiles.
We see a Gaussian like profile in the lower part of the hopper (see data at $z=60~d^*$) and a profile 
with a clear plateau and narrow shear zones at the two sides at a higher elevation ($z=150~d^*$).
The shape difference is clearly captured by the exponent $\eta$ (see Fig.~\ref{fig:velocity}(c))
which is between 2-3 for $z=60~d^*$ and around 9 for $z=150~d^*$.

%
%

Turning to the case of rods with $L/d=6$ we see, that velocity fluctuations are so large, that they
are already affecting the shape of the velocity profile. Figure \ref{fig:collapse}(c) shows one of the
most violent events, when the velocity profile collapses and the flow almost stops. We see, that the
velocity profile before collapse has a clear plateau, and sharp steps (very narrow shear zones) at
the two sides. During collapse however, its shape changes significantly.  Such shape changes lead
to the fact, that the steps on the two sides of the time averaged velocity profile (shown with a
dashed line in Fig.~\ref{fig:collapse}(c)) became less steep, resulting in a smaller value of the
exponent $\eta$ (about 5.6 in this case) than for rods with $L/d=1.4$.  This effect was clearly
noticeable for rods with $L/d=3.5$ and $L/d=6$.

In a way, this strange behavior of long grains is in qualitative agreement with observations of long
cylindrical particles in 3D silos. There, we have identified so-called ''rat holes'' which form above the
orifice when the aspect ratio of the particles becomes larger than six \cite{ashour2017}. Those holes represent vertical
tunnels with stable side walls above the orifice, where the material remains stagnant at the sides of the
rat hole, while the silo empties only by the material inside the rat hole. The silo discharge stops
even without clogging when the rat hole penetrates the granular bed in the silo and reaches the surface.
In the 2D experiments, we see essentially the same feature that the material remains stagnant at the
sides and flow is restricted to a kind of two-dimensional rat hole above the orifice. The outflow does not
come to a complete rest, and violent avalanches can destroy part of the stagnant zones temporarily.


\section{Summary}

We have experimentally studied the flow field of a granular material in a quasi-two-dimensional
hopper. Using six granular samples with different grain shapes (spherical, oblate and prolate), we find
that the velocity profile $-$ characterizing the downward motion of the grains $-$ can be well
fitted with a Gaussian function for spherical particles as earlier models predicted, however
for elongated grains the flow field has a different form. In that case the flowing region is
narrower and is bordered with sharper velocity gradient. We quantified the deviation of the
velocity profile from the Gaussian form by measuring the exponent $\eta$ as a function of
the vertical position in the hopper. Focusing on the time evolution of the velocity profile,
we find that the flow of elongated grains is characterized by velocity fluctuations of larger
amplitude and lower frequency compared to the case of spheres.
\\

\section*{ACKNOWLEDGMENTS}
This work was supported by the Hungarian National Research, Development and Innovation
Office NKFIH under grant OTKA K 116036 and by the DAAD/M\"OB researcher exchange program
(Grants No. 29480 and 64975).

\end{document}